\documentclass[aip, apl, amsmath,amssymb, reprint]{revtex4-1}

\usepackage{graphicx}
\usepackage{dcolumn}
\usepackage{bm}
\usepackage[usenames]{color}

\def\uband{13.5$\pm$1\%\;nm\;}
\def\eband{11.2$\pm$1\%\;nm\;}

\begin{document}
\title[]{Observation of extreme-ultraviolet light emission from an expanding plasma jet with multiply-charged argon or xenon ions}
\author{A. G. Shalashov}
\email[Author to whom correspondence should be addressed. Electronic mail:\;]{ags@appl.sci-nnov.ru}
\author{A. V. Vodopyanov}
\author{I. S. Abramov}
\author{A. V. Sidorov}
\author{E. D. Gospodchikov}
\author{S. V. Razin}
\affiliation{Institute of Applied Physics RAS, 603950 Nizhny Novgorod, Russia}
\author{N. I. Chkhalo}
\author{N. N. Salashchenko}
\affiliation{Institute for Physics of Microstructures RAS, 603950 Nizhny Novgorod, Russia}
\author{M. Yu. Glyavin}\author{S. V. Golubev}
\affiliation{Institute of Applied Physics RAS, 603950 Nizhny Novgorod, Russia}
\date{\today}

\begin{abstract}

We report the first direct demonstration of possibility to generate the extreme ultraviolet (EUV) radiation with a freely expanding jet of dense plasma with multiply-charged ions supported by high-power microwaves. The detected emission power is about 20 W at 18--50\,nm for argon and xenon and 0.3 W at 13--17\,nm for xenon. The discharge with a peak electron density up to $3\times 10^{16}\,$cm$^{-3}$ and a characteristic size of $150\,\mu$m is supported by a focused radiation of a recently developed gyrotron with unique characteristics, 250~kW output power at 250~GHz, operating in a relatively long (50$\,\mu$s) pulse mode.
Up-scaling of these experimental results gives grounds for development of a point-like kilowatt-level EUV source for a high-resolution lithography able to fulfill requirements of the microelectronics industry.
\end{abstract}

\maketitle

Next generation high-resolution projection lithography for chip production requires a powerful and reliable source of radiation at the short-wavelength boundary of the extreme ultraviolet (EUV) range\cite{bakshi_2006,wagner_2010,moore}. The only practical generation mechanism in this range is the line radiation of elements in a highly ionized state, such as Sn$^{7+}$--Sn$^{12+}$ that have a significant number of strong radiation lines at \uband \cite{churilov_2003}, and Xe$^{10+}$ that has a family of strong lines at \eband \cite{churilov_2002,churilov_2004,saloman_2004}. These two spectral bands are essentially eminent as they both correspond to peak reflection coefficients of available multilayer mirrors used for manipulation with EUV light \cite{ckhalo_adv_2013}. 

While all existing prototypes of an industrial EUV source operate with laser-produced Sn-plasma \cite{giga_mizoguchi_2017,asml_schafgans_2015}, 
a microwave discharge provides an alternative potentially less constrained in terms of output EUV power, characterized with a simpler overall design and safe operation for EUV focusing optics which is free from being spoiled by solid target particles and fast ions typical of explosive (of a few nanoseconds) laser-produced plasmas. Microwave pulses last longer, from tens of microseconds to continuous-wave operation, and there are no channels for significant ion heating due to direct resonant power load into electrons. In pioneer experiments, a strongly non-equilibrium Sn plasma was confined in an open magnetic trap and heated by radiation of a high-power 75 GHz / 50 kW gyrotron, resulting in up to 50 W emission in \uband band into 4$\pi$ sr \cite{Vodopyanov,ChkhVod,Vodopyanov1}. Theoretical modeling shows good potentials of this scheme for the development of an industry-ready EUV source featured with a multi-kW level in \uband band as most required hardware, namely compact high-current sources of tin plasma and high-power gyrotrons, are available \cite{abramov_pop_2017}.

A microwave discharge can naturally be used to produce EUV emission from heavy noble gases, an option which is also being studied for laser-produced plasmas 
\cite{ckhalo_2018}. For the spectral band of \eband, featured of line-radiation of xenon, there are Ru/Be and Mo/Be mirrors that have nearly two-fold higher peak reflection coefficient in comparison to Mo/Si mirrors in \uband band \cite{ckhalo_adv_2013}. Thus, a combination of xenon as an emitter and new mirrors potentially is more attractive for EUV lithography development \cite{abramov_pra_2018}.
The first experiment aimed at the realization of a point-like highly-emissive discharge in a noble gas has been reported \cite{glyavin_apl_2014}. 
Such experiment becomes possible with the development of a gyrotron providing 100 kW output power at 670 GHz frequency in 20 $\mu$s pulse\cite{subTHz1}. Going to higher frequencies allows supporting a microwave discharge at higher plasma densities ($\sim 10^{16}\;$cm$^{-3}$), and more narrow focusing of the radiation. At this, no external confining magnetic field is needed---the discharge is ignited in a target gas jet launched from a small nozzle at almost the atmospheric pressure, and then freely expanding into a pumped-out chamber (7--200\,mTorr at the wall). In these experiments, the possibility of formation of a point-like hot plasma spot near the nozzle was proved, and a total emission up to 10 kW in 110--180\,nm band was detected from argon plasma. 

In this letter, we continue the experiment with the freely expanding plasma. The experiment is arranged around a new gyrotron with unique characteristics of 250~kW at 250~GHz \cite{subTHz2,subTHz3}. Although this tube is designed for CW operating, due to limitations of the available power supply presently it works in a pulsed mode, the pulse length up to 50$\,\mu$s with the repetition rate 10\,Hz. Using a higher microwave power level and longer pulses, allow us to investigate in detail the dynamics of the plasma emission and to perform the experiment in xenon. Our main result is a detection of plasma emission in the EUV spectral ranges of 13--17 nm (for xenon) and 18--50 nm (for xenon and argon). 

\begin{figure}[t]
\includegraphics[width=80mm]{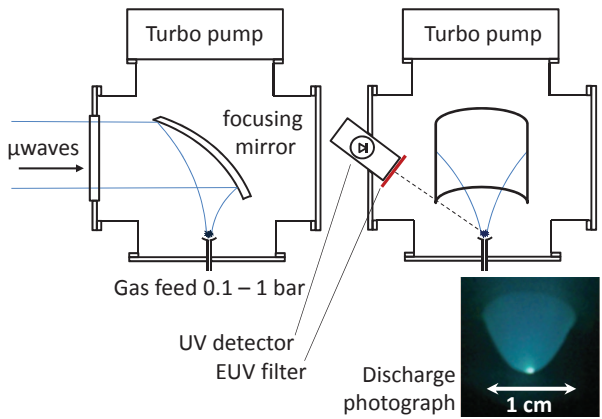}
\caption{The schematic of the experiment (front and side views) and a photograph of the point-like discharge.}\label{Fig01}
\end{figure}

The schematic of the experiment is presented in Fig.~\ref{Fig01}. Neutral gas is injected into a vacuum chamber with a nozzle. The nozzle diameter is 150$\,\mu$m, the pressure behind the nozzle is varied from 0.1 to 1 bar. High-power microwave radiation from a gyrotron is focused in front of the nozzle and is resonantly absorbed by electrons under conditions of the plasma resonance when the electron Langmuir frequency becomes equal to the wave frequency. This condition is locally met somewhere along the jet with a decreasing plasma density; a high electron thermal conductivity provides an efficient heat transport towards more dense plasma near the nozzle \cite{shalashov_jetp_2016}. The direct power load into electrons leads to formation of nonequilibrium high-electron-temperature plasmas with cold ions, 
characterized by high rates of the electron impact ionization/excitation and a suppressed recombination. It is beneficial for generation of highly charged ions able of emitting in EUV band. 

The background pressure in the chamber is kept at the level below few mTorr to make media transparent for the EUV light. The decrease of plasma density with the jet expansion provides a good localization of the discharge resulting, in particular, in a point-like EUV emitting region near the nozzle, less than 1 mm in all dimensions.
 Noting that the optimal pressure for a microwave breakdown at 250 GHz is close to 200 Torr, the breakdown conditions are fulfilled only in the small area near the nozzle \cite{sidorov_smp_2017}, where the plasma density rises above $10^{16}\,$cm$^{-3}$ at the developed stage of a discharge \cite{sidorov_pop_2016}.

\begin{figure}[tb]
\includegraphics[width=80mm]{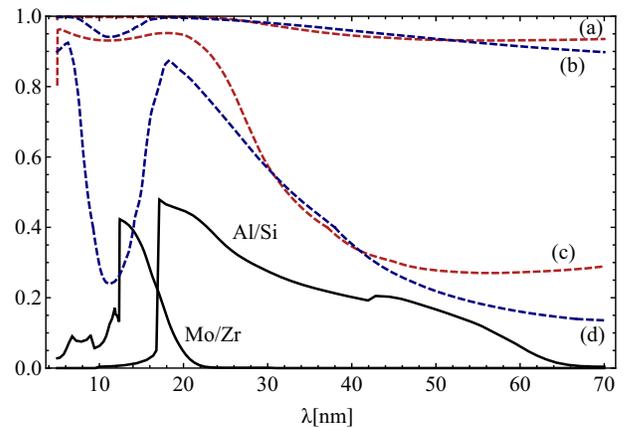}
\caption{ Transmission coefficients of EUV filters $C_{{f}}(\lambda)$ (solid lines) and neutral gas $C_n(\lambda)$ (dashed lines) for different linear densities $\eta_n$: (a) Ar at $1.5\times 10^{15}$cm$^{-2}$, (b) Xe at $1.5\times 10^{15}$cm$^{-2}$, (c) Ar at $4.5\times 10^{16}$cm$^{-2}$, (d) Xe at $6\times 10^{16}$cm$^{-2}$. }\label{Fig02}
\end{figure}

An absolutely calibrated silicon detector with a set of filters is used to measure ultraviolet light properties \cite{V-det}. We use Mo/Zr filter for the 13--17 nm band and Al/Si filter for the 18--50 nm band \cite{V-fil}; spectral characteristics of the filters are shown in Fig.~\ref{Fig02}. To avoid errors due to accidental holes both filters are used in single and double combinations. The detector has an effective radius $R=0.3\,$cm and is placed at $L=40\,$cm from the nozzle at different angles to the plasma. The received signals are independent of the observation angle. Assuming an isotropic emission with a flat spectrum inside a filter bandwidth $\Delta\lambda$, one may link the detected power $P_\mathrm{det}$ with a total power $P$ of plasma emission in $4\pi$ sr as
$$P_\mathrm{det}=\left(C_d\int C_f(\lambda)\,\frac{\mathrm{d}\lambda}{\Delta\lambda}\right)P,\quad C_d=\frac{\pi R^2}{4\pi L^2},$$
where $C_d$ is the detector function, $C_f$ is the filter characteristic.

\begin{figure*}[bt]
\includegraphics[width=173mm]{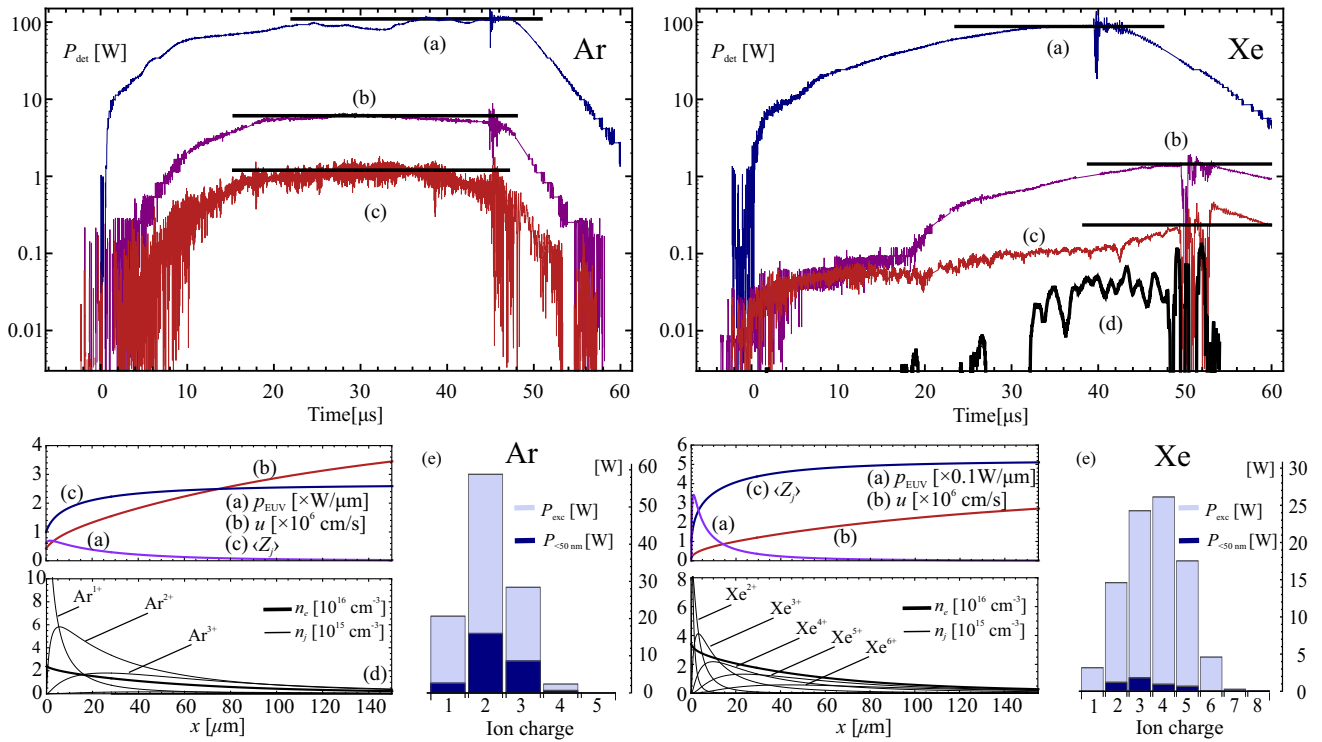} 
\caption{Top: measured oscillograms of UV luminosity of argon and xenon point-like plasma with different filters: (a) no filter, (b) 18--50\,nm single, (c) 18--50\,nm double, (d) 13--17\,nm single. Microwave pulses last from 0 to 40--50$\,\mu$s, the heating power is 180 kW for argon and 250 kW for xenon. 
Bottom: (a)--(d) calculated discharge parameters as a function of coordinate $x$ along the flow (distance from the nozzle output): (a) linear density $p_{\text{EUV}}$ of the power losses in the spectral band of 10--50 nm, (b) flow velocity $u$, (c) average ion charge $\langle Z_j\rangle$, (d) electron density $n_e$ (thick line) and densities of ion species $n_j$ (thin lines); (e) distribution of the emission power (total and in 10--50 nm band) over ion charges. Other parameters are listed in Table \ref{Tab01}. Free parameters of the numerical simulations are found by fitting the reference levels of measured signals indicated with horizontal lines in the top plots.
}\label{Fig03}
\end{figure*}

\begin{table}[bt]
\caption{\label{Tab01} Measured and calculated parameters of the EUV-emitting point-like discharge. Simulations are performed by fitting $F$, $T_e$, $\eta_n $ to the experimental signals.}
\begin{center}
\begin{tabular}{p{0.26\textwidth}p{0.095\textwidth}p{0.095\textwidth}}
\hline 
Parameter& Argon & Xenon$\vphantom{\big)^A}$ \\ \hline
\multicolumn{3}{ c }{\textit{Experiment}$\vphantom{\big)^A}$}\\ 
EUV power, 13--17 nm & -- & 0.3 W\\
EUV power, 18--50 nm & 22 W & 5.5 W\\
Conversion efficiency, 18--50 nm & $1.2\!\cdot\! 10^{-4}$ & $3\!\cdot\! 10^{-5}$ \\
Total emitted power, no filter & 110 W & 90 W\\
Microwave power & 180 kW & 250 kW \\
Optimal pressure behind nozzle & 0.55 bar & 0.4 bar \\ \hline
\multicolumn{3}{ c }{\textit{Simulations}$\vphantom{\big)^A}$}\\ 
Plasma flux $F$ & $1.8\!\cdot\! 10^{18}$s$^{-1}$ & $0.9\!\cdot\! 10^{18}$s$^{-1}$\\
Electron temperature $T_e$ & 45 eV & 45 eV \\
Linear density of halo neutrals $\eta_n$ & $4.5\!\cdot\! 10^{16}$cm$\!\!^{-2}$ & $6\!\cdot\! 10^{16}$cm$\!^{-2}$ \\
Maximal electron density $n_e $ & $2.4\!\cdot\! 10^{16}\rm{cm}\!\!^{-3}$ & $3.4\!\cdot\! 10^{16}\rm{cm}\!\!^{-3}$ \\
Maximal average ion charge $\langle Z_j \rangle$ & 2.7 & 5.2 \\
\hline
\end{tabular}
\end{center}
\end{table}

Typical signals of the point-like plasma emission measured with and without EUV filters during the microwave pulse are presented in Fig.~\ref{Fig03}. Although the optimal conditions for a maximal luminosity are essentially the same in argon and xenon, the signals show a slightly different evolution in time.

For argon, an initial growth during the plasma set-up time (necessary to establish a high-charge state) is followed by a stationary phase, and then a decay after the gyrotron is switched-off. Key parameters of the stationary phase are summarized in Table \ref{Tab01}. The maximum EUV power of plasma emission at 18--50 nm, $P= 22\,$W, is achieved with the heating power of 180 kW and the pressure behind the nozzle of 0.55 bar. Emission of argon plasma at 13--17\,nm is almost zero. The most part of the light belongs to the deep UV range. 

For xenon, there is no pronounced stationary phase: the luminosity increases with time reaching a maximum value at the end of the heating pulse. Data in Table \ref{Tab01} correspond to times just before the gyrotron switch off. EUV emission from xenon plasma is observed in both available spectral bands. In the 13--17 nm band, the maximal light power of 0.3 W is achieved with the heating power of 250 kW and the nozzle pressure of 0.4 bar. At these parameters, the light power at 18--50 nm is 5.5 W. The experimental error in measuring the power is primarily due to a pulse-to-pulse repeatability: a jitter of few microseconds in plasma appearance leads to about 10--15\% deviation of the luminosity.

Comparing to our previous experiments\cite{glyavin_apl_2014}, one can see a dramatic, by a factor of 400, degradation of the conversion efficiency of the heating radiation into the ultraviolet light \cite{fn}. 
This may be explained by not optimal focusing of the heating microwave beam at a lower frequency $\omega_{\rm{mw}}$, at which the size of the plasma formation is much less than the beam waist. 
On a qualitative level one may consider a multiplication of two effects: the microwave power at the focal point scales as $\omega_{\rm{mw}}^2$, the coupling to a small spherical plasmoid scales approximately as the Rayleigh scattering cross-section $\sigma_R\sim d^6\omega_{\rm{mw}}^4$. In combination, we obtain that the efficiency of microwave absorption scales as $\omega_{\rm{mw}}^6$, what corresponds to a factor of 370 for 670~GHz/250~GHz. More rigorous approach to the calculation of microwave absorption on an inhomogeneous plasma sphere \cite{IEEE} predicts an essentially the same degradation of the coupling efficiency 
(by chance, the experiment at 670~GHz hits the optimal conditions for the plasma heating). 

Direct measurements of most key parameters, such as electron temperature and plasma density, for a compact non-stationary strongly inhomogeneous and non-equilibrium discharge are not available in our experiment. Thus, for the interpretation of measurements, we strongly rely on a numerical model developed specifically for the experiment \cite{abramov_pra_2018}. The model is based on quasi-one-dimensional fluid equations which are solved together with kinetic equations for step-by-step ionization and line-excitation, including EUV radiation transport in an optically-thick plasma. Assuming the particular law of jet expansion, only two free input parameters remain: the electron temperature $T_e$ and the total (conserved) plasma flux $F$. Then, we can determine distributions along the flux of all essential discharge features, such as the directed flow velocity $u$, the densities of ions with different charges $n_j$, the electron density $n_e$, the density of power losses for line-excitation and emission $p_\mathrm{exc}$, etc. 
So, these parameters may be estimated by adjusting only $T_e$ and $F$ to fit the simulations to the signals shown in Fig.~\ref{Fig03}. 
The signal coming to our detector is simulated with taking into account the instrumental function $C(\lambda)$ of the receiver and filters as
\begin{equation}\label{eq1}
P_{\mathrm{det}}=\int \sum_{j}\sum_{h,l} C(\lambda_{jhl}) \Delta E_{jhl} k_{jlh}^* n_e n_{j} \;\mathrm{d}V,
\end{equation}
where 
indexes $h$ and $l$ numerate energy levels of excited and ground electron configuration for all allowed transitions in the spectrum of $j$-th ion, $\lambda_{jhl}$ and $\Delta E_{jhl}$ are the transition wavelength and energy, $k_{jlh}^*$ is the \emph{effective} line excitation coefficient introduced to take into account the possibility of radiation trapping inside the emitting volume \cite{abramov_pra_2018}, and $C=C_d C_f C_n$ with $C_d$ and $C_f$ being the detector function and the filter characteristic, $C_n$ is explained further. Taking $C=1$ one obtains exactly the formula for total plasma emission losses, $P_{\mathrm{exc}}$, see Eq.~(2) in Ref.~\onlinecite{abramov_pra_2018}.

However, applying this strategy we face two difficulties. First, our model describes only a stationary phase of a discharge, a non-stationary extension is still a target of our efforts. So, we have no problem in the interpretation of data for argon; but for xenon, we must assume voluntary that the discharge is close enough to the steady-state before the gyrotron is switched off. 

Second problem is that before a modification the model systematically predicted a higher relative level of EUV light compared to deep UV light. This is solved by considering an absorption of ultraviolet light in a residual neutral gas. In previous researches, this factor has been neglected assuming a good coupling of the microwave radiation to plasma that results in an effective ionization of the whole jet. Since in the present experiment the coupling efficiency is far from being optimal, one may expect a skin of low ionized matter surrounding the emitting discharge. Photo-ionization of this matter by UV quanta may be taken into account by adding an additional transmission factor $C_n$, defined similar to a photon escape probability $\theta_{jhl}$ in Ref.~\onlinecite{abramov_pra_2018}: 
$$C_n(\lambda_{jhl})=\!\int\! a(\omega)\,\mathrm{e}^{-\tau a(\omega)/a(0)}\,\mathrm{d}\omega, 
\;\; \tau=\sigma_\nu(\lambda_{jhl})\,\langle n_n r_\nu\rangle,
$$
where $a(\omega)$ is the Voigt line profile function, i.e.\ a convolution of the Doppler and natural lines normalized to a frequency integral of unity, $\omega$ is a detuning from the central line, $\sigma_\nu(\lambda_{jhl})$ is the photo-ionization cross-section\cite{kennedy}, $n_n$ is the density of neutrals, $r_\nu$ is the linear distance that a photon should pass to escape in vaccuum, and $\langle ...\rangle$ denotes the averaging over a jet volume. Note that $\eta_n=\langle n_n r_\nu\rangle$ is a new additional parameter used in the fitting; it characterizes an average density of neutrals in the halo. 
In calculations, we use approximate formulas for a photon escape probability developed by Apruzese\cite{apruzese_1985}. The resulted transmission factors as a function of wavelength are shown in Fig.~\ref{Fig02}. 
For each gas, we present two curves: one is for characteristic $\eta_n$ at which $C_n>0.9$ at all wavelengths, e.g.\ the neutral gas becomes transparent (curves \textit{a}, \textit{b}), the other one is corresponding to an actual $\eta_n$ obtained by fitting to the experimental data (\textit{c}, \textit{d}).


The absorption in the halo may indeed depress the EUV spectra at longer wavelengths, although for xenon there is an additional non-transparent window at $\lambda\sim10\,$nm. The latter means that an effective EUV light source with xenon would eventually imply very strong demands to the quality of vacuum inside a working chamber. Most of the absorbed power is finally re-emitted during recombination. The minimal wavelengths of recombination radiation defined by the ionization potential are 79 nm for argon and 103 nm for xenon. Having this in mind, we take into account $C_n$ only when calculating the EUV power at $\lambda<50\,$nm, and omit it when calculating the total emission power detected without filters. 


Figure \ref{Fig03} and Table \ref{Tab01} show the results of simulations fitted to the most strong set of signals measured with 18--50 mn filter. 
We reconstruct the main discharge parameters as varying functions along the flow and the distribution of radiation power losses over the ion charges. Fitted electron temperature is of $45\,$eV. Calculated electron density $\sim 3\times10^{16}\,\rm{cm}^{-3}$ is in good agreement with the results of measurements in similar conditions \cite{sidorov_pop_2016}. More heavy xenon is more easily ionized by an electron impact than argon, therefore it is featured with higher electron densities and average ion charge. However, the xenon ions are still not ionized enough to emit effectively at \eband. 
A fairly high effective density of neutrals in the halo, $\langle n_n\rangle\sim\eta_n/d\sim(3-4)\times10^{18}\rm{cm}^{-3}$ where $d\sim 150\,\mu$m is a characteristic transverse size of the jet, comparable to the density of neutrals inside the nozzle, leaves some potentials for essential improvement of the EUV efficiency with a better chamber pump-out and conditioning.

In summary, we have shown the possibility of direct conversion of high-power microwaves into the EUV light in the ionized jet of noble gases. An advantage of such approach for a high-resolution projection lithography is a relatively simple design, a compactness of the point-like emitting zone, and a possibility of CW operation with available microwave sources. The main drawback of the existing experiment is a low conversion efficiency related to a poor coupling of the heating wave and point-like plasma dictated by the low frequency of available gyrotron.
Further increase in the conversion efficiency can be achieved either by use a more sophisticated electrodynamic system (resonator), or by increasing the frequency of the heating wave in the existing scheme. 



\vspace{6pt}
The authors are grateful to all colleagues from the Department of Electronic Devices for the support of gyrotron operation. 
The work is supported by the Russian Science Foundation (grant No. 14-12-00609).


\begin{thebibliography}{6}

\bibitem{bakshi_2006}
V. Bakshi, \emph{EUV Sources for lithography} (The Society of Photo-Optical Instrumentation Engineers, 2006).

\bibitem{wagner_2010}
C. Wagner, N. Harned, Nature Photonics \textbf{4}, 24 (2010).

\bibitem{moore} S. K. Moore, IEEE Spectrum \textbf{55} (1), 46 (2018).

\bibitem{churilov_2003}
S. Churilov, Y. N. Joshi, J. Reader,  Optics Letters \textbf{28}, 1478 (2003).

\bibitem{churilov_2002}
S. S. Churilov, Y. N. Joshi, Physica Scripta, \textbf{65}(1), 40 (2002).

\bibitem{churilov_2004}
S. S. Churilov, Y. N. Joshi, J. Reader, R. R. Kildiyarova,  Physica Scripta \textbf{70}(2-3), 126 (2004).

\bibitem{saloman_2004}
E. B. Saloman, Journal of Physical and Chemical Reference Data \textbf{33}(3), 765 (2004).

\bibitem{ckhalo_adv_2013}
N. I. Chkhalo, N. N. Salashchenko, AIP Advances, \textbf{3(8)}, 082130 (2013).


\bibitem{giga_mizoguchi_2017}
H. Mizoguchi, H. Nakarai, T. Abe, K. M. Nowak, Y. Kawasuji, H. Tanaka, Yu. Watanabe, T. Hori, T. Kodama, Yu. Shiraishi, T. Yanagida, T. Yamada, T. Yamazaki ,S. Okazaki, and T. Saitou, Proc. of SPIE \textbf{10143}, 101431J-1 (2017).

\bibitem{asml_schafgans_2015}
A. A. Schafgans, D. Brown, I. Fomenkov, R. Sandstrom, A. Ershov, G. Vaschenko, R. Rafac, M. Purvis, S. Rokitski, Ye. Tao, D. Riggs,
W. Dunstan, M. Graham, N. Farrar, D. Brandt, N. Bowering, A. Pirati, N. Harned, C. Wagner, H. Meiling, R. Kool, Proc. SPIE \textbf{9422}, 94220B-1 (2015).


\bibitem{Vodopyanov} 
A. V. Vodopyanov, S. V. Golubev, D. A. Mansfeld, A. G. Nikolaev, K. P. Savkin, N. N. Salashchenko, N. I. Chkhalo, and G. Yu. Yushkov, JETP Lett. \textbf{88}, 95 (2008).

\bibitem{ChkhVod} 
N. I. Chkhalo, S. V. Golubev, D. Mansfeld, N. N. Salashchenko, L. A. Sjmaenok, A. V. Vodopyanov, Journal of Micro/Nanolithography, MEMS, and MOEMS \textbf{11}, 021123-1 (2012)

\bibitem{Vodopyanov1} 
A. V. Vodopyanov, S. V. Golubev, D. A. Mansfeld, N. N. Salashchenko, and N. I. Chkhalo, Bull. Russ. Acad. Sci.: Phys. \textbf{75}, 64 (2011)

\bibitem{abramov_pop_2017}
I. S. Abramov, E. D. Gospodchikov, A. G. Shalashov, Physics of Plasmas \textbf{24}(7), 073511 (2017).

\bibitem{ckhalo_2018} N. I. Chkhalo, S. A. Garakhin, S. V. Golubev, A. Ya. Lopatin, A. N. Nechay, A. E. Pestov, N. N. Salashchenko, M. N. Toropov, N. N. Tsybin, A. V. Vodopyanov,  S. Yulin, Appl. Phys. Lett. \textbf{112}, 221101 (2018).

\bibitem{abramov_pra_2018}
I. S. Abramov, E. D. Gospodchikov, A. G. Shalashov, 
Extreme-ultraviolet light source for lithography based on an expanding jet of dense xenon plasma supported by microwaves,
arXiv:1712.10026 [physics.plasm-ph].

\bibitem{glyavin_apl_2014} 
M. Y. Glyavin, S. V. Golubev, I. V. Izotov, A. G. Litvak, A. G. Luchinin, S. V. Razin, A. V. Sidorov, V. A. Skalyga, A. V. Vodopyanov, Applied Physics Letters \textbf{105}(17), 174101 (2014).

\bibitem{subTHz1} M. Yu. Glyavin, A. G. Luchinin, G. S. Nusinovich, J. Rodgers, D. G. Kashyn, C. A. Romero-Talamas, and R. Pu, Appl. Phys. Lett. \textbf{101}, 153503 (2012).
 
\bibitem{subTHz2} M. Glyavin, G. Denisov, Proc. of IRMMW-THz 2017 42nd International Conference, 1 (2017).

\bibitem{subTHz3} G. G. Denisov, M. Yu. Glyavin, A. P. Fokin, A. N. Kuftin, A. I. Tsvetkov, A. S. Sedov, E. A. Soluyanova, M. V. Bakulin, E. V. Sokolov, E. M. Tai, M.V.Morozkin, M.D.Proyavin, V.E.Zapevalov.
First experimental tests of powerful 250GHz gyrotron for future fusion research and CTS diagnostics, Rev.Sci.Instr. (2018, in print)



\bibitem{shalashov_jetp_2016}
A. G. Shalashov, I. S. Abramov, S. V. Golubev, E. D. Gospodchikov, JETP \textbf{123}, 219 (2016).

\bibitem{sidorov_smp_2017} 
A. V. Sidorov, S. V. Razin, A. G. Luchinin, A. I. Tsvetkov, A. P. Fokin, D. S. Sidorov, A. P. Veselov, A. V. Vodopyanov, M. Y. Glyavin, EPJ Web of Conferences \textbf{149}, 02031 (2017).

\bibitem{sidorov_pop_2016}
A. V. Sidorov, S. V. Razin, S. V. Golubev, M. I. Safronova, A. P. Fokin, A. G. Luchinin, A. V. Vodopyanov,  M. Yu. Glyavin, Phys. Plasmas \textbf{23} (4), 043511 (2016).

\bibitem{V-det} 
P. N. Aruev, M. M. Barysheva, B. Ya. Ber, N. V. Zabrodskaya, V. V. Zabrodskii, A. Ya. Lopatin, A. E. Pestov, M. V. Petrenko, V. N. Polkovnikov, N. N. Salashchenko, V. L. Sukhanov, N. I. Chkhalo,
Quantum Electronics \textbf{42} (10), 943 (2012) [DOI:10.1070/QE2012v042n10ABEH014901].

\bibitem{V-fil} 
A. D. Akhsakhalyan, E. B. Kluenkov, A. Y. Lopatin, V. I. Luchin, A. N. Nechay, A. E. Pestov, V. N. Polkovnikov, N. N. Salashchenko, M. V. Svechnikov, M. N. Toropov,  N. N. Tsybin, N. I. Chkhalo, A. V. Shcherbakov,
Journal of Surface Investigation: X-ray, Synchrotron and Neutron Technique \textbf{11}(1), 1 (2017) [DOI:10.1134/S1027451017010049]. 





\bibitem{fn} 
The microwave power focused into plasma at 670~GHz is 40\,kW, it results in the 10\,kW UV emission (with 25\% efficiency); the same procedure at 250~GHz gives the efficiency of $110\,\rm{W}/180\,\rm{kW}=6\times10^{-4}$.

\bibitem{IEEE} A. Shalashov, E. Gospodchikov,  IEEE Transactions on Antennas and Propagation  \textbf{64} (9), 3960 (2016); JETP \textbf{123} (4), 587 (2016).

\bibitem{kennedy} D. J. Kennedy, S. T. Manson, Phys. Rev. A \textbf{5}(1), 229 (1972).

\bibitem{apruzese_1985}
J. P. Apruzese,   Journal of Quantitative Spectroscopy and Radiative Transfer \textbf{34}(5), 447 (1985).


\end{thebibliography}
\end{document}